\documentclass{vgtc}                          

\ifpdf
  \pdfoutput=1\relax                   
  \usepackage{graphicx}                
  \DeclareGraphicsExtensions{.pdf,.png,.jpg,.jpeg} 
\else
  \usepackage{graphicx}                
  \DeclareGraphicsExtensions{.eps}     
\fi%

\graphicspath{{figures/}{pictures/}{images/}{./}} 

\usepackage{microtype}                 
\PassOptionsToPackage{warn}{textcomp}  
\usepackage{textcomp}                  
\usepackage{mathptmx}                  
\usepackage{times}                     
\usepackage{cite}                      
\usepackage{tabu}                      
\usepackage{booktabs}                  
\usepackage{enumitem}   
\usepackage{xcolor}

\onlineid{0}

\vgtccategory{Research}

\vgtcinsertpkg

\newcommand{\MyColorBox}[2][red]%
{%
    \settowidth{\Width}{#2}%
    \colorbox{#1}%
    {%
        \raisebox{-\DepthReference}%
        {%
                \parbox[b][\HeightReference+\DepthReference][c]{\Width}{\centering#2}%
        }%
    }%
}

\title{Literal Encoding: Text is a first-class data encoding}

\author{Richard Brath\thanks{e-mail: rbrath@uncharted.software}\\ %
        \scriptsize Uncharted Software Inc.} %

\abstract{Digital humanities are rooted in text analysis. However, most visualization paradigms use only categoric, ordered or quantitative data. Literal text must be considered a base data type to encode into visualizations. Literal text offers functional, perceptual, cognitive, semantic and operational benefits. These are briefly illustrated with a subset of sample visualizations focused on semantic word sequences, indicating benefits over standard graphs, maps, treemaps, bar charts and narrative layouts.%
} 

\CCScatlist{
  \CCScatTwelve{Human-centered computing}{Visu\-al\-iza\-tion}{Visu\-al\-iza\-tion techniques};
  \CCScatTwelve{Human-centered computing}{Visu\-al\-iza\-tion}{Visualization design and evaluation methods}{}
}

\begin{document}


\firstsection{Introduction}

\maketitle

The role of text in visualization is poorly defined. In most reference works on visualization the focus is on categoric, ordered or quantitative data as base types of data that are transformed into visual attributes then drawn in a display, e.g. \cite{Bertin1967, Cleveland1984, Mackinlay1986, Wilkinson1999, Mazza2009, ChenFloridi2013 }. Text is usually understood as categoric data such as labels added on a graph or words in a tag cloud. However, categoric, ordered and quantitative encoding alone do not consider the unique role of literally encoding text. This paper will show functional, perceptual, cognitive, semantic and operational benefits that literal text affords visualizations.

\begin{figure}[tb]
 \centering 
 \includegraphics[width=\columnwidth, trim={0 6cm 0 3cm},clip]{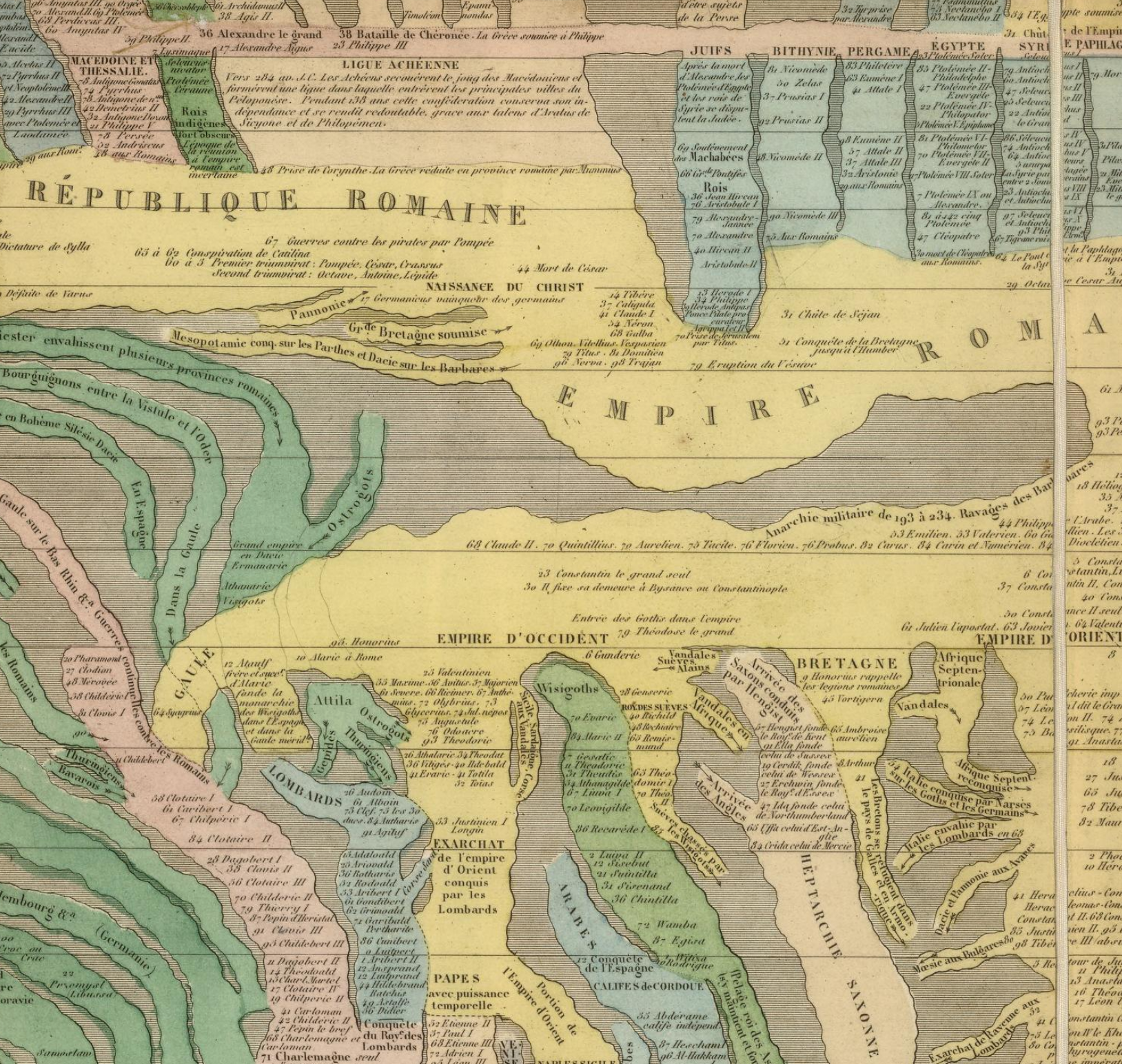}
 \caption{Text-intensive chronological visualization of world history.}
 \label{fig:tableau}
\end{figure}

Consider the portion of the world history chart in \autoref{fig:tableau} \cite{Pic58}: it is a time-oriented visualization showing empires as streams: essentially a storyline visualization (e.g. \cite{TM12} based on xkcd.com/657). However, it also has a large amount of text including the names of empires in big text, rulers, dates, key events, short descriptive phrases, full sentences and so on: the informational content far exceeds storyline visualizations and the encodings are beyond categoric data.

\section{BACKGROUND}
Literal text goes beyond categories or individual words.

\subsection{More than categories}
Categoric data in a visualization is typically transformed into a visual attribute such as hue or shape. However, perception limits the number of unique hues to ten or fewer \cite{Ware2000}. While it is feasible to have an unlimited number of unique shapes \cite{Bertin1967}, in practice very few unique shapes are provided by default in off-the-shelf software and programming libraries (e.g. nine in Excel, ten Tableau, seven D3.js). 

There are tens of thousands of unique English words, with thousands understood by most adult speakers. There is a larger set of proper nouns. Then highly unique word sequences can be constructed to form phrases, sentences, stories, poetry and so forth. Further, words are much richer than categories and can express nuance – e.g. many alternate words exist for each emotion \cite{Plu01}.

\subsection{More than individual words}
Combinations of words, relationships between words and sequences of words are typically lost in popular visualization approaches. Tag clouds typically represent individual words, e.g. \cite{Feinberg2010}. Simple chart types, such as bar charts and pie charts, often use labels which require short text and tools often truncate text automatically. Some visualization techniques do work with longer text sequences such as WordTrees \cite{WV08}, Arc Diagrams \cite{Wat02} or some of the knowledge maps at scimaps.org (e.g. \cite{BHH07}) – meaning that visualization of longer sequences do exist, but, these are not common.

Closely related shortened strings are the underlying analytic techniques that reduce text to individual words or phrases such as word frequency analysis, entity detection, sentiment analysis, emotion analysis, and vector-based representations such as Word2Vec (e.g. \cite{MCCD13}). As such, the analysis of text by separating words loses much semantic context: Working with text in visualization means reconsidering text as a base type of data.

\subsection{History}
There are many historic visualization examples of text-based representations. Even within prose, highlighting techniques such as changes in color, italics, capitalization and weight have been used for centuries to make letters or words perceptually standout from surrounding text. And there are many other examples of text heavy visualizations such as: 

\begin{figure}[htb]
 \centering 
 \includegraphics[width=\columnwidth, trim={0 2cm 0 0},clip]{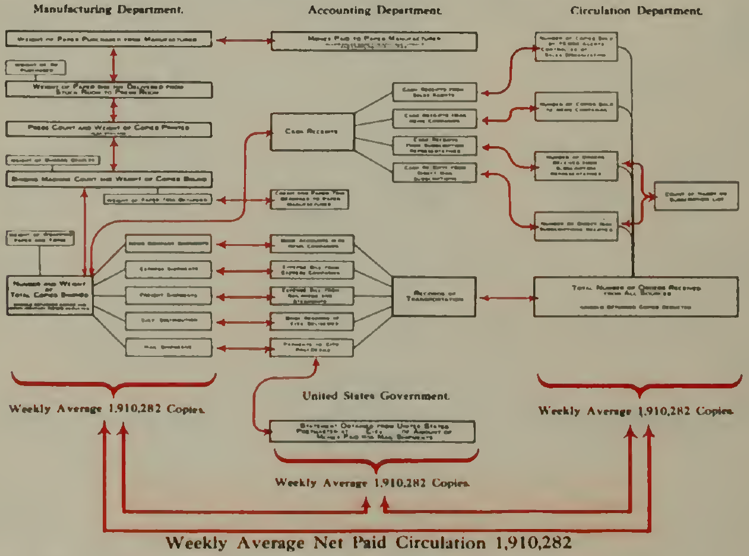}
 \caption{Portion of a text-intensive flowchart.}
 \label{fig:flowchart}
\end{figure}

\textit{Flowcharts} document processes and may need text to describe each step, such as in \autoref{fig:flowchart} (from \cite{Brinton1939}).

\textit{Timelines} may be heavily annotated such as Chapple and Garofalo’s \textit{History of Rock N Roll} (in \cite{Tufte1990}); may contain names and phrases such as Figure 1; detailed sentences such as Rand McNally’s \textit{Histomap} \cite{Spa31}, or many other examples in \textit{Cartographies of Time} \cite{RG12}).

\begin{figure}[htb]
 \centering 
 \includegraphics[width=\columnwidth, trim={0 11cm 0 0},clip]{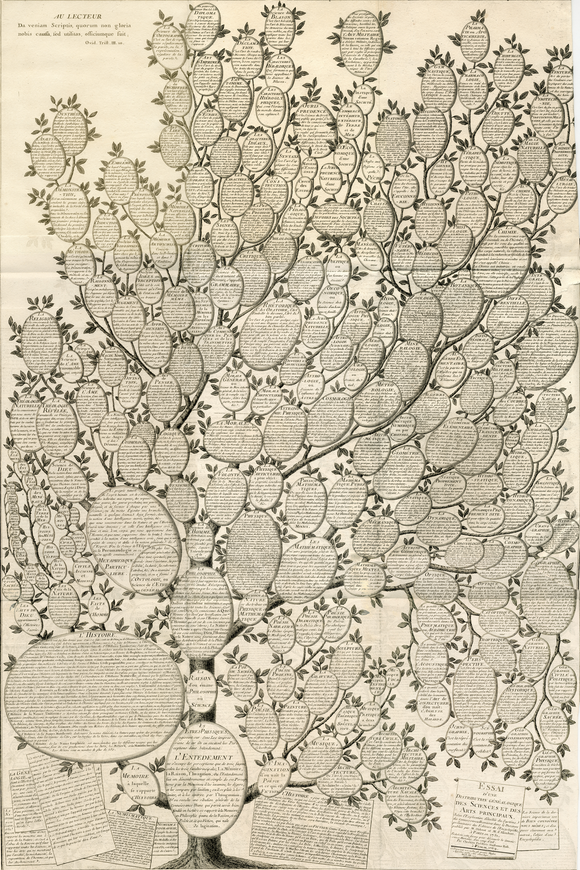}
 \caption{Top part of Diderot's \textit{Tree of Knowledge}.}
 \label{fig:diderot}
\end{figure}

\textit{Knowledge maps }and \textit{mind maps} require text, some of which contain large amounts of descriptive text such as Diderot’s \textit{Tree of Knowledge} in \autoref{fig:diderot} \cite{Rot80}. Modern procedurally-generated knowledge maps are text oriented as well, such as the proportional Euler diagrams in openknowledgemaps.org \cite{KKE16}.

\subsection{Information Visualization}
In the field of information visualization, there are more than 400 text visualizations at the \textit{Text Visualization Browser} (textvis.lnu.se). However, approximately one quarter of these visualizations have no text at all. Most of the remaining techniques are focused on labels, i.e. a word or two. Only a few manipulate individual characters (such as background per glyph \cite{ARL13} or font weight \cite{NHC12}). Some display longer text, although many of these limit visualization to manipulation of keywords in context (KWIC, e.g. \cite{Hearst2009}) and/or blocks of text; e.g. \cite{SKSK18} does both.

Use of literal text, beyond simple labels, may be uncommon in information visualizations for many reasons \cite{Brath2020}, such as: 

\begin{enumerate}[topsep=3pt,itemsep=0pt,partopsep=0pt, parsep=0pt]
    \item \textit{Historic convention} separates words from images due to limitations of print-based processes or the extra expense associated with engravings, lithographs, etc.
    \item \textit{Display limitations} in early visualizations resulted in poor quality text (low resolution 96DPI) and limited real-estate (small screen sizes), which prompted guidelines to display details on demand \cite{Shn96}.
   \item \textit{Graphic design conventions}, starting with mid-century modernism, favor less text, more white space and more graphics such as icons.
   \item \textit{Preattentive pattern perception} precludes reading text linearly.
\end{enumerate}

These issues can be addressed, for example:
\begin{enumerate}[topsep=3pt,itemsep=0pt,partopsep=0pt, parsep=0pt]
    \item Modern web browsers do not have the same restrictions as print.
    \item Modern displays are much higher resolution: a 4k display has 25x pixels of an early 1990's display.
    \item Graphic design counter-trends bring more text manipulation into media, starting with post-modernists in the 1980's.
   \item Some visualization tasks do not require preattention (e.g. inventories such as roadmaps), or subsets of text can be made preattentive with typographic formats \cite{Bertin1967, bertin1980, Brath2020}.
\end{enumerate}


\subsection{Characterization and Examples}
The rest of this paper, and the key contribution, will characterize the benefits of literal text directly encoded in visualization and illustrate some examples focused on visualizations of strings of text.

\section{Literal Encoding}
Data can be encoded literally into strings of text and represented directly. Literal encodings are unique to text. Instead of considering text as a different form of data, some researchers consider text as a visual attribute, similar to shape, size or color, e.g. \cite{Wilkinson1999, Mazza2009, ChenFloridi2013, Borner2015Atlas}, whereas other researchers do not include text. Bertin, for example, does not include text as a data type or encoding, however, Bertin narrowly defined his visual attributes as retinal variables, explicitly focusing on low level visual channels into which data is transformed for thematic representations for fast perception. Note that Bertin did discuss text and font attributes such as bold, italic and typeface in four pages in an appendix to the original French edition of Sémiologie Graphique \cite{Bertin1967} and a followon article \cite{bertin1980}, which were not included in the later English translation.

\begin{figure}[tb]
 \centering 
 \includegraphics[width=\columnwidth]{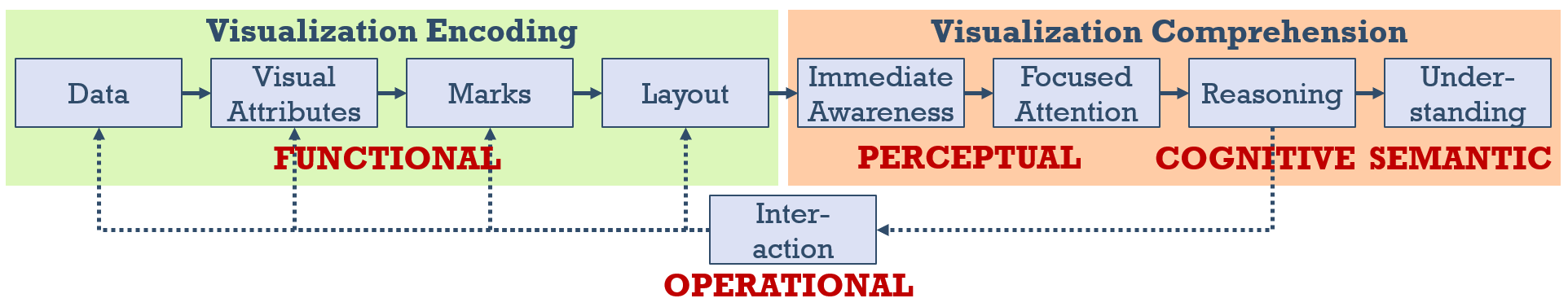}
 \caption{Literal text benefits (red text) in relation to the visualization pipeline.}
 \label{fig:pipeline}
\end{figure}

There are many benefits to literal encoding, which can be organized by function, perception, cognition, semantics and operation. These benefits can be aligned to the \textit{visualization pipeline} as shown in \autoref{fig:pipeline} (this diagram based on a simplified version from \cite{ChenFloridi2013}).

\subsection{Functional Benefits} Literal text may be effective in meeting the needs associated with the particular use case:

\textbf{A. Information organization or communication purpose.} The goal of a visualization is not necessarily analysis of patterns by preattentive perception, such as a pattern of dots in a scatterplot. Bertin defines other purposes, including communicating key information and organizing a large amount of data. Text is common in historic visualizations that organize large amounts of information as shown in the prior section. It is doubtful that these data-dense visualizations which organize information could function without text.

\textbf{B. Text is the primary subject.} Some visualizations are primarily about text. The simple tag cloud could not exist without text. Similarly, other visualizations related to text analysis presumably will have a strong need to explicitly represent literal text: extracted entities, topic analysis, sentiment and emotion analytics, keyword analysis, word stemming, dialogue analysis, taxonomies, and so on.

\textbf{C. Increased information content.} Many historic visualizations are highly similar to modern interactive visualizations. For example, Figure 1 is a storyline visualization. Flowcharts like Figure 2 are essentially graphs with much larger nodes sized to fit descriptive text. Fundamentally, visualization is a lossy medium \cite{ChenFloridi2013}, and the addition of text into visualization can increase the informational content.

\textbf{D. High number of categories.} Many visual attributes do not support categories of high cardinality as discussed in 2.1. Text can support high cardinality as words, phrases and sentences; or codes e.g. labeling aircraft in air traffic control, equipment in network diagrams, nodes in circuit diagrams, and so on.

\textbf{E. Reduced design effort.} It is feasible pictographic icons can be quick to decipher, but effective icons typically require significant effort by expert graphic designers to create a series of icons which work together.

\textbf{F. Unambiguous Decoding.} While an icon can represent a word or phrase, they can be ambiguous, e.g. Clarus the dogcow \cite{Hac15}. Text literally encodes the specific words, phrases or sentences of interest, so there is no information loss decoding back to text. Further, textual glyphs are highly learned and unambiguous in a well-designed font.

\subsection{Perceptual Benefits: Fast Identification}

Many visualization evaluations focus on testing time and errors, such as the rapid perception of the presence of a visual attribute. While text can be enhanced with attributes such as bold or italic to make a word preattentively pop-out, to read the text still requires active attention - which is slower than preattention.

However, real-world user tasks may be much more complex, requiring full decoding and comprehension of context. If the goal of a visualization is overall more efficient completion of tasks, then the performance of the entire system must be considered. Active attention of text can provide multiple perceptual benefits:

\textbf{A. Reading is Automatic.} Reading labels is very fast compared to interactions such as tooltips. The parallel letter recognition model of words states letter information is used to recognize the words \cite{larson2004}, letters within a word are recognized simultaneously, and the and word perception is extremely fast \cite{Dav03}.

\textit{Automatic word recognition} is a common explanation for the Stroop effect which theorizes that reading is \textit{automatic} and difficult to voluntarily stop. The \textit{relative speed of processing model}, hypothesizes that it is faster to read the word than to name the color. Both explanations imply that reading words is fast. More generally, automaticity is the ability to perform a well practiced task with low attentional requirements: (a) the user is unaware that the task is occurring, (b) does not need to initiate the task, (c) does not have the ability to stop the process, and, (d) the task has low cognitive load \cite{Bar94}. The implication for visualization is that some text will be automatically read and that the cognitive cost of reading will be low.

\textbf{B. Interaction is Slow.} Interactions such as tooltips can be used to identify items on a screen. Interaction requires additional cognitive effort, as the viewer must first determine their interactive goal, engage in motor skills and progressively refine those skills to achieve the target \cite{PV07}. 

Zoom and pan with level of detail for progressive appearance of labels is an alternative approach to revealing text, often used on interactive maps. While zoom-based labeling does occur in visualization (e.g. \cite{Jonker2017}), it is not common. Visualization does not have the equivalent labeling heuristics developed over centuries in cartography \cite{RMM95}. 

\textbf{C. Legends are slow.} In any form of visualization the viewer must both perceive the information of interest and then decode it. When visually assessing categoric data encoded as discrete instances in a visual attribute (e.g. using color to indicate categories), the viewer needs to recall the mapping between the colors and categories. With text, the viewer can directly decode the item.

\textbf{D. Reduced load on short term memory.} Tasks reliant on short-term human memory can benefit from explicit text. Text represented directly does not require short term memory, for example when referring back and forth between items and legends.

\subsection{Cognitve Benefits: Aid to Reasoning}

Beyond attention, problem-solving tasks require more cognitive resources to complete more complex tasks, such as determining the segment in a Venn diagram that corresponds to a logical condition, or, tracing a path in a network. Complex reasoning can be aided by the appropriate visual representations, as shown with force diagrams, pulley combinations and so forth (e.g. see many of the papers at \textit{Diagrams} conference). 

Readers construct mental representations of what they read at multiple levels: a) surface: words and syntax; b) propositional content; c) situation model incorporating and organizing content with respect to real world knowledge from memory \cite{Pay07}.

\textbf{A. Proposition and problem-solving tasks.} The propositional model is self contained based on the presented facts: it can be understood through logical inferences, or through spatialization and visually queried. Problem solving is improved when textual instructions are directly integrated into diagrams \cite{CS91}. Task performance improves when information is provided through both text and images with close spatial positioning and/or linkages \cite{NH02}. More broadly, Larkin and Simon’s findings indicate that diagrammatic representations of content aid problem-solving \cite{LS87} by:
\begin{enumerate}[topsep=3pt,itemsep=3pt,partopsep=0pt, parsep=0pt]
    \item \textit{Locality:} Spatially organizing information together reduces search effort.
    \item \textit{Reducing cross-referencing: }Fewer steps to decode mappings.
   \item \textit{Perceptual enhancement:} Visual relationships support perceptual inferences.
\end{enumerate}

\begin{figure}[tb]
 \centering 
 \includegraphics[width=\columnwidth]{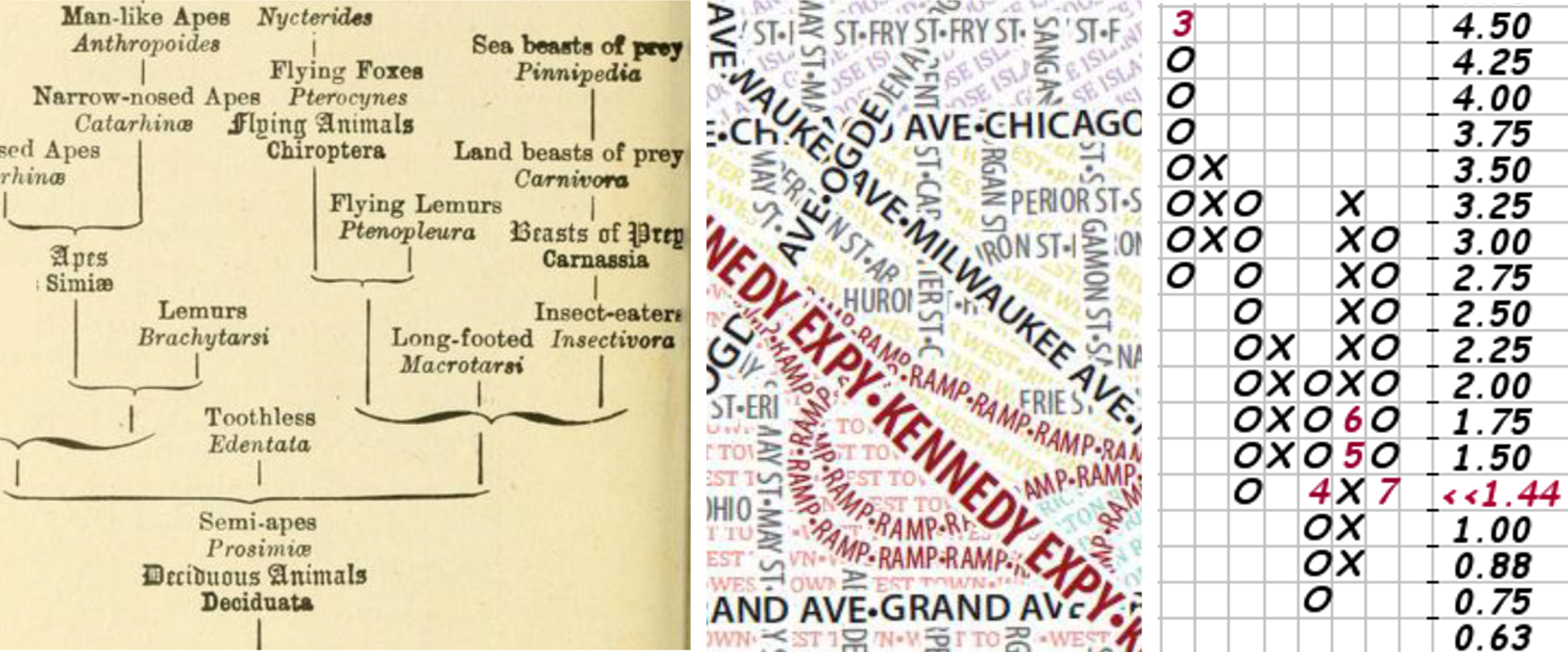}
 \caption{Perceptual inferences can be made via text density, lines and areas, or patterns forming columns.}
 \label{fig:haeckl}
\end{figure}

Extending Larkin and Simon’s findings to text and visualization implies improved performance for text directly integrated into a visualization:

\begin{enumerate}[topsep=3pt,itemsep=3pt,partopsep=0pt, parsep=0pt]
    \item \textit{Reduced Search (Locality). }Spatially organized displays aid finding information of interest. For example, in the knowledge graph (e.g. \autoref{fig:diderot}), the hierarchical relations and descriptions are spatially grouped facilitating search and providing text details. Identification of local relations is a common task in many visualizations and has been expressed by many researchers, for example:
    
        \paragraph{Tobler's first law of geography} “Everything is related to everything else, but near things are more related than distant things.” \cite{Mil04} Many visualization layouts attempt to locate related objects close together (e.g. graph layouts, scatterplots, treemaps and set visualization). Text in any of these layouts supports identification and context of local related entities.
    
        \paragraph{Thundt et al's serendipity} Serendipitous discovery is the fortuitous unexpected discovery by accident, often occurring during search. It is researched in library sciences and summarized by Thudt et al \cite{thudt2012bohemian} who note that serendipity is closely associated with coincidence: wherein related ideas may manifest as simultaneous occurrences that seem acausal but meaningful. Both local proximity and related words and phrases can be recognized thereby providing different cognitive associations than visual associations alone.

    \item \textit{Reduced Cross-Referencing}. Text within a visualization element provides immediate detail as opposed to drill-down or referring to other linked visualizations: 
    
    \paragraph{Tufte, Shneiderman and details} Tufte popularized micro/macro readings; that is visual displays with high-density information, visually read at levels ranging from high-level macro patterns such as trends, clustering and outliers; down to localized patterns, such as individual observations and local peers \cite{Tufte1983}. Tufte summarizes this with “to clarify, add detail.” Tufte’s approach is somewhat similar to Shneiderman’s visual information-seeking mantra: “overview first, zoom and filter, then details on demand.” \cite{Shn96} With Shneiderman, interactivity is used to reveal the detail information whereas Tufte plots it directly. Tufte plotted data on high-resolution paper, whereas Shneiderman was constrained to low-res 1990’s screens, necessitating interaction.
    
    \paragraph{Bertin and categories} Bertin states that shapes can represent categories of high cardinality and provides a map with 59 different glyph types (p. 157 in \cite{Bertin1967}). Bertin shows global patterns with respect to one shape cannot be seen. The viewer must linearly scan across glyphs and can only see local patterns. Furthermore, the viewer cannot ascertain if a local pattern is meaningful without comparison to the global pattern. Text can be different. Text can be used non-categorically to encode unique entities and concepts (e.g. countries, people, idioms, quotes, etc.) – a comparison to global is not needed. Furthermore, any unique text can be accessed non-linearly based on the layout, for example, to directly compare the unique text at two different locations of the plot area. 

    \item \textit{Perceptual inferences} can be made across collections of labels or phrases. For example, a density of text can indicate common nodes across classification schemes in the graph by Haeckel (\autoref{fig:haeckl} left, \cite{Haeckel97}), the alignment of text along paths to form lines and areas (e.g. Axis’ Maps: Typographic Maps in \autoref{fig:haeckl} middle \cite{Afzal2012}), or the repetition of X’s and O’s in point and figure charts (\autoref{fig:haeckl} right), e.g. \cite{DeVilliers1933}. 
    
\end{enumerate}

\subsection{Semantic Benefits: Word Sequences}

Semantic content is added to text when words are combined in sequences. Consider the following:

\begin{itemize}[topsep=3pt,itemsep=0pt,partopsep=0pt, parsep=0pt, leftmargin=36pt]
    \item	Jack and Jill went up the hill.
    \item   Humpty Dumpty sat on a wall.
   \item    The Owl and the Pussycat went to sea.
\end{itemize}

Text, when processed into discrete words, such as a word cloud, loses the semantic content of the word sequence. These sentences have meaning beyond the collection of words. Jack, Jill and Humpty all have height and might fall. Jack, Jill, the Owl and the Cat went somewhere, Humpty did not.

Analytic and visualization approaches need to consider applications where word sequence is maintained, or assemblies of related words. Search user interfaces have evolved significantly beyond simple keywords and document metadata to include contextual titles, phrases and sentences in search results (e.g. \cite{Hearst2009} Chapter 5). Studies have shown superior performance for results containing text snippets \cite{CAD07}. 

More broadly, a variety of visualization techniques have evolved in the humanities for close reading, summarized by Jänicke et al \cite{JFC15}. For close reading, these approaches maintain word sequence and change visual attributes to markup the text (similar to keyword markup), superimpose markers and connections on top of words (e.g. like a graph), or increase space between letters and words to add markers such as flows or phonetics (e.g. like WordTree).

\subsection{Operational Benefits: Text Manipulation}
Literal text allows for textual operations which can be used to pre-process text, or interactively to operate on text to provide for additional uses:
\begin{itemize}[topsep=3pt,itemsep=3pt,partopsep=0pt, parsep=0pt]
    \item \textit{Ordering.} Text can be ordered and sorted (alphabetically), facilitating search and lookup.
    \item \textit{Search and Filter.} Text can be navigated and/or filtered by complete or partial strings.
    \item	\textit{Summarization.} Text can be summarized: longer texts can be reduced to shorter texts while retaining a subset of the original meaning.
    \item \textit{Comparison, similarity and translation.} Text can be compared and assessments made regarding similarity. Tools such as thesauruses and dictionaries aid in understanding similarity. Translation extends comparative analysis to produce near equivalent meaning in another language.
    \item \textit{Tone, opinion, sentiment and emotion.} Text can be evaluated assigned quantitative values, for example, for opinion (e.g. ratings), tone (e.g. news tone), sentiment (e.g. score for positive or negative), or emotion (e.g. based on text or other cues).
    \item \textit{Categorization, taxonomies and topic analysis.} Organization of many texts follow classification schemes and tagging of topics by keywords; such as general purpose classifications (e.g. Dewey, Library of Congress) and domain specific classifications (e.g. ACM, IEEE).
    \item \textit{Natural Language Processing and Machine Learning.} All of the above operations are rapidly evolving with NLP and machine learning, such as topic extraction, emotion analysis, sentiment detection, machine translation, automated summarization and so on. For example, Zhang et al.'s recent transformer models have achieved near human-level abstractive summarization \cite{zhang2020pegasus}. Open source NLP libraries, such as NLTK, spaCy, and compromise \cite{Nltk,Spacy,Compromise}, provide parts of speech tagging, tokenization, entity extraction, dependency parsing, and entity linking, which can be assembled into a wide variety of textual analyses extracting and linking related words and phrases, such as such as character descriptions or rhetorical devices.  
\end{itemize}

\section{Discussion}
There are many ways to consider these benefits in the humanities. As there are many existing visualization techniques focused on collections of disconnected words (e.g. word clouds), the following focuses on applications relevant to semantic word sequences (i.e. phrases and sentences) and the corresponding benefits.

\begin{figure}[htb]
 \centering 
 \includegraphics[width=7cm]{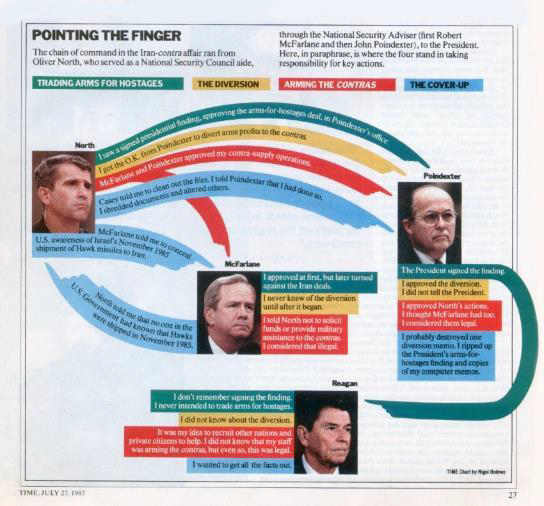}
 \caption{Connections between individuals with testimony.}
 \label{fig:holmes}
\end{figure}

\textbf{A. Sentences on Paths.} In \autoref{fig:holmes}, Nigel Holmes reduced key testimony in the Iran-Contra affair into a set of statements implicating individuals represented as color-coded textual connections between politicians (i.e. a graph) \cite{Holmes1987}. Literal text is required for the communications purpose (3.1:A). Magazines are non-interactive – the text is a more efficient alternative to referring to a legend (3.2:C), and aids reasoning with directed connections enabling perceptual inferences (3.3:A3). 

\begin{figure}[tb]
 \centering 
 \includegraphics[width=\columnwidth]{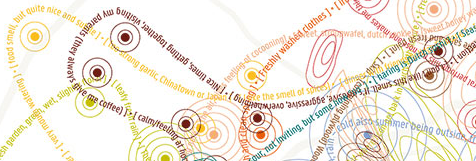}
 \caption{Portion of Amsterdam SmellMap with narrative paths.}
 \label{fig:mclean}
\end{figure}

Kate McLean uses linear text along geographic contour lines to form artistic geospatial narratives in \autoref{fig:mclean} \cite{McLean2014}. The literal text is the primary subject (3.1:B), which the viewer will automatically start to read (3.2:A), and compare local information within a geolocation (3.3:A1), with added potential for residents to supplement real-world local knowledge (3.3:B).

Programmatic examples include news headlines with associated stock price movement, and twitter commentary with associated retweet counts \cite{Brath2020}; which enable comparison between content (the text) and the response (the behaviour of markets / sharing by users).

\begin{figure}[htb]
 \centering 
 \includegraphics[width=\columnwidth]{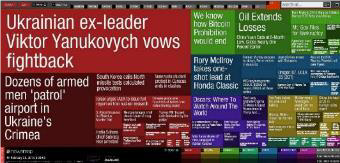}
 \caption{NewsMap headlines. Size, hue and brightness encode data.}
 \label{fig:weskamp}
\end{figure}

\vspace{5pt}
\textbf{B. Small Blocks of Text.} \textit{NewsMap} extends the visualization layout of a treemap with news headlines \autoref{fig:weskamp} \cite{Weskamp2004}. Information content beyond a treemap is increased with hue, brightness and text (3.1:C). Interaction is not a suitable: direct reading is required (3.2:B). The treemap layout of text ensures that similar headlines are spatially proximate (3.3:A1) and augments the viewer’s real-world understanding of news (3.3:B). Interactive search and filter aid exploration (3.5). 

One challenge with the treemap layout is the prioritization of area accuracy over the display of text, as such, some squares are too small for legibile text, or text does not fit well. Consider a dataset of 2894 coroners inquests from Georgian London \cite{HSH12,How18} where each case is summarized by a sentence and verdict, e.g. 

\begin{itemize}[topsep=3pt,itemsep=0pt,partopsep=0pt, parsep=0pt]
 \item Mary Roberts drowned herself. Suicide.
 \item Mary Gardiner struck with hand. Homicide.
 \item Ann Fitsall suffocated and burnt. Accident.
 \item Nicholas Bone, John Dayson and James Cusack killed by a brick wall. Accident.
\end{itemize}

\begin{figure}[htb]
 \centering 
 \includegraphics[width=\columnwidth]{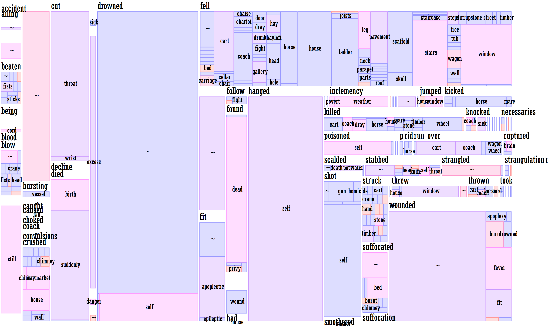}
 \caption{Treemap showing causes of death in Georgian London.}
 \label{fig:treemap}
\end{figure}

An analysis may be interested in causes of death. From each case can be extracted a subject, verb and object. These can be organized into a hierarchy, by verb (e.g. drowning, suffocation), object (e.g. hand, wall) and subject (e.g. Mary Roberts, Ann Fitsall). There are a few caveats, for example, where two or more verbs or objects are used in a single sentence, only one is selected so that individuals are not double counted. The hierarchy can be visualized, for example, with a treemap (\autoref{fig:treemap}), which draws attention to big boxes and bright colors (e.g. drowned) but at the cost of fitting labels or skipping labels on small boxes.

\begin{figure}[tb]
 \centering 
 \includegraphics[width=\columnwidth]{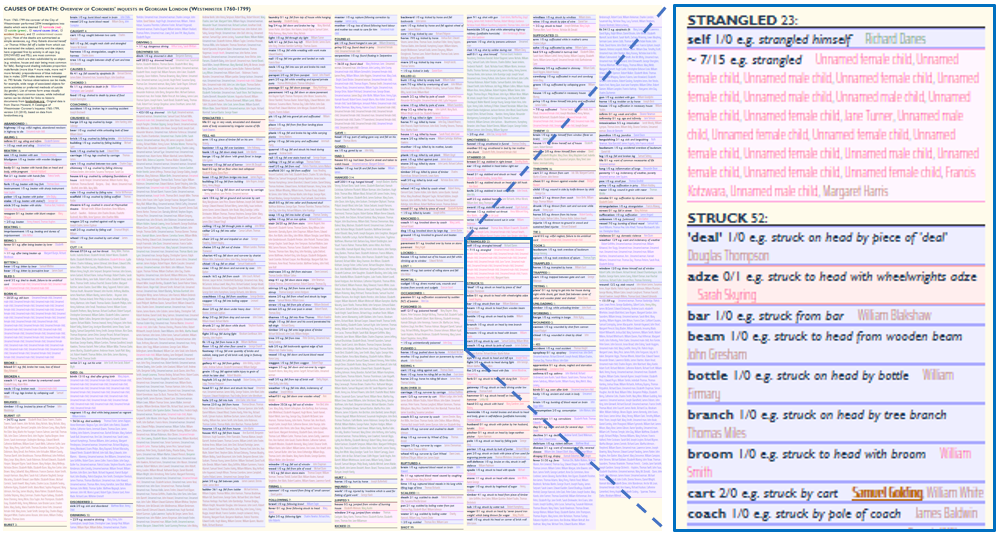}
 \caption{Activities, objects, and the names of deceased.}
 \label{fig:london}
\end{figure}

Instead, \autoref{fig:london} shows the same data in a representation conceptually similar to a dictionary listing with font weight, color and background shading conveying data, live at \url{https://codepen.io/Rbrath/full/rNewgde}. All text is visible and readable on a 4K display: an enlarged portion is shown at right (i.e. organizational purpose 3.1:A). Broad patterns are visible, for example, large blocks of names such as the red text (homicide) on red background (gender female) under the verb \textbf{STRANGLED} (i.e. a preattentive perception of the large block, then the bold allcaps text, which is identified by automatic reading 3.2:A). On close inspection, the local text can be read, for example, many local objects associated with the verb \textbf{STRUCK} (e.g. \textbf{adze, bar, beam, bottle, broom}) and the names of the victims (e.g. \textbf{adze}: \textcolor{magenta}{Sarah Skyring}, \textbf{bar}: \textcolor{brown}{William Blakshaw}, etc) – no drill-down is needed (3.3:A2). Furthermore, search can be used to find the common forms of death for “child”, or filters can be used to isolate only homicides (3.5). 

\begin{figure}[tb]
 \centering 
 \includegraphics[width=\columnwidth]{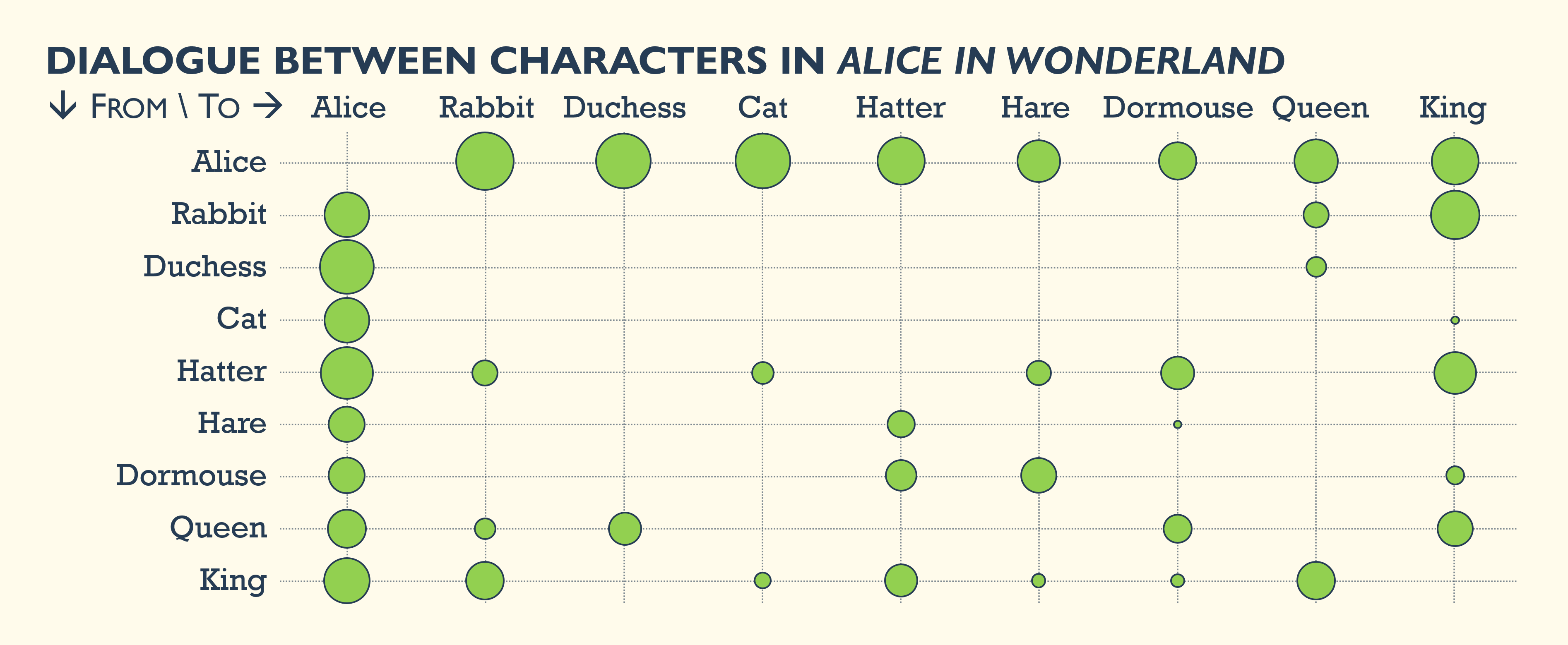}
 \caption{Volume of character dialogue via an adjacency matrix.}
 \label{fig:adjmat}
\end{figure}

\autoref{fig:adjmat}, is a simple adjacency matrix indicating dialogue from one character (vertical axis) to another character (horizontal axis), with bubble size proportional to the amount of dialogue.

\begin{figure}[tb]
 \centering 
 \includegraphics[width=\columnwidth]{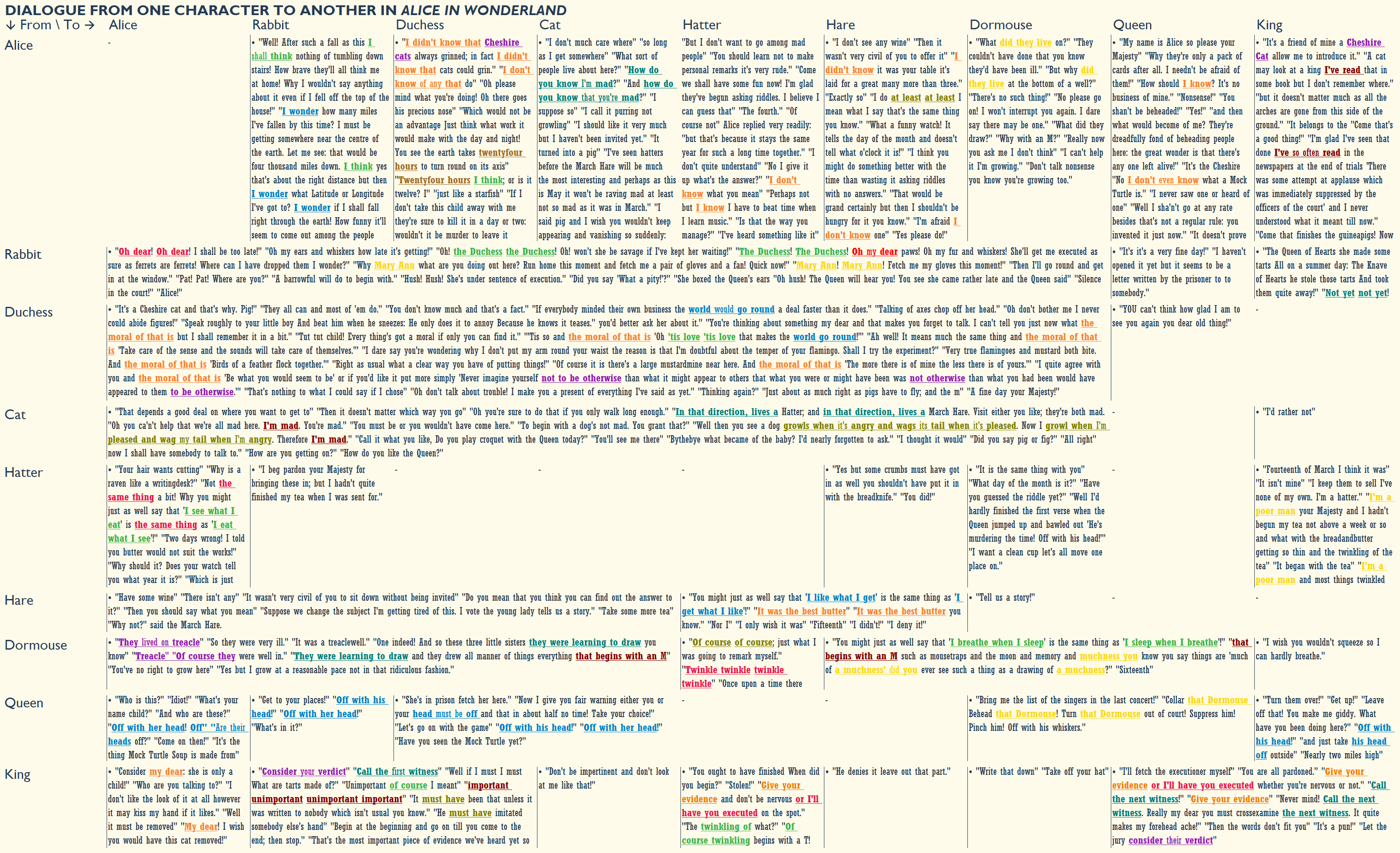}
 \caption{Character dialogue via an adjacency matrix, with highlighted repeated sets of words per character.}
 \label{fig:adjtext}
\end{figure}

 	\definecolor{dkgreen}{RGB}{20,100,0}
 	\definecolor{dkcyan}{RGB}{0,80,100}
\autoref{fig:adjtext} is the same adjacency matrix where each cell contains the dialogue from one character to another, increasing content (3.1:C). Cells are expanded if the next cell to the right would otherwise be empty. The matrix contains much of the dialogue, although some cells are truncated. In addition to showing dialogue, some phrases have been highlighted: an simple NLTK algorithm has tagged sets of words that are frequently repeated by characters (not including sets made of only prepositions, articles and conjunctions), so for example, \textcolor{orange}{“the moral of that is”} is highlighted each time the Duchess repeats the phrase to Alice (aiding perceptual inferences 3.3:C). Word sets are repeated words, but the order may vary. This also highlights Carroll’s logical inversions, such as the Hatter’s \textcolor{dkgreen}{“I see what I eat”} and \textcolor{dkgreen}{“I eat what I see”}. It also highlights variants of catchphrases, such as \textcolor{dkcyan}{“off with his/her head”} yelled by the Queen. Highlighting similar words in different orders aids analysis of the variation of the semantic content (3.4). This use of NLP could be further tuned, for example, to tag rhetorical devices; which in turn, could be visualized in techniques such as an adjacency matrix, story flow, text on path, and so on. 

\vspace{5pt}
\textbf{C. Superimposition.} Text can be superimposed over top of data-driven graphical marks – similar to using a highlighter, where the highlight may represent quantities or keywords or so on.

\begin{figure}[htb]
 \centering 
 \includegraphics[width=\columnwidth]{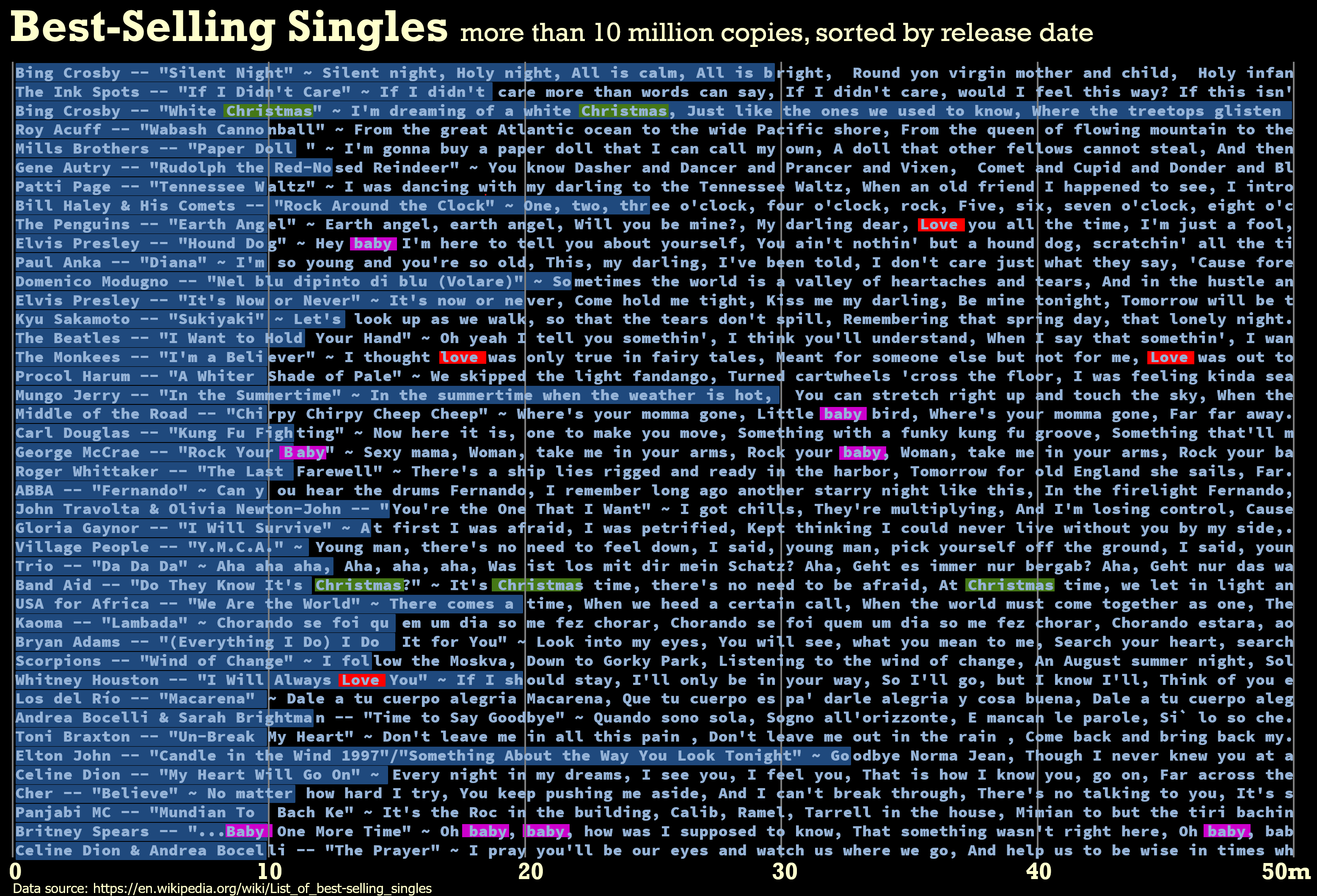}
 \caption{Song title and lyrics over bar chart and repeated words.}
 \label{fig:singles}
\end{figure}

 	\definecolor{ltblue}{RGB}{127,127,255}
Each row in \autoref{fig:singles} is a top selling song, shown as a line of text identifying artist, song tile, and opening lyrics. The line is superimposed over a \colorbox{ltblue}{blue bar chart} indicating number of singles sold and also superimposed over highlights of common keywords such as \colorbox{green}{Christmas}, \colorbox{red}{love}, and \colorbox{magenta}{baby}. The superimposed text per bar increases information beyond the standard bar chart (3.1:C); provides detail instead of an interactive tooltip (3.2:B), allows perceptual inferences (patterns of repeated words) (3.3:A3) and reading can trigger real world recall of a familiar tune (3.3:B). 

\begin{figure}[htb]
 \centering 
 \includegraphics[width=\columnwidth, trim={0 7.5cm 0 0},clip]{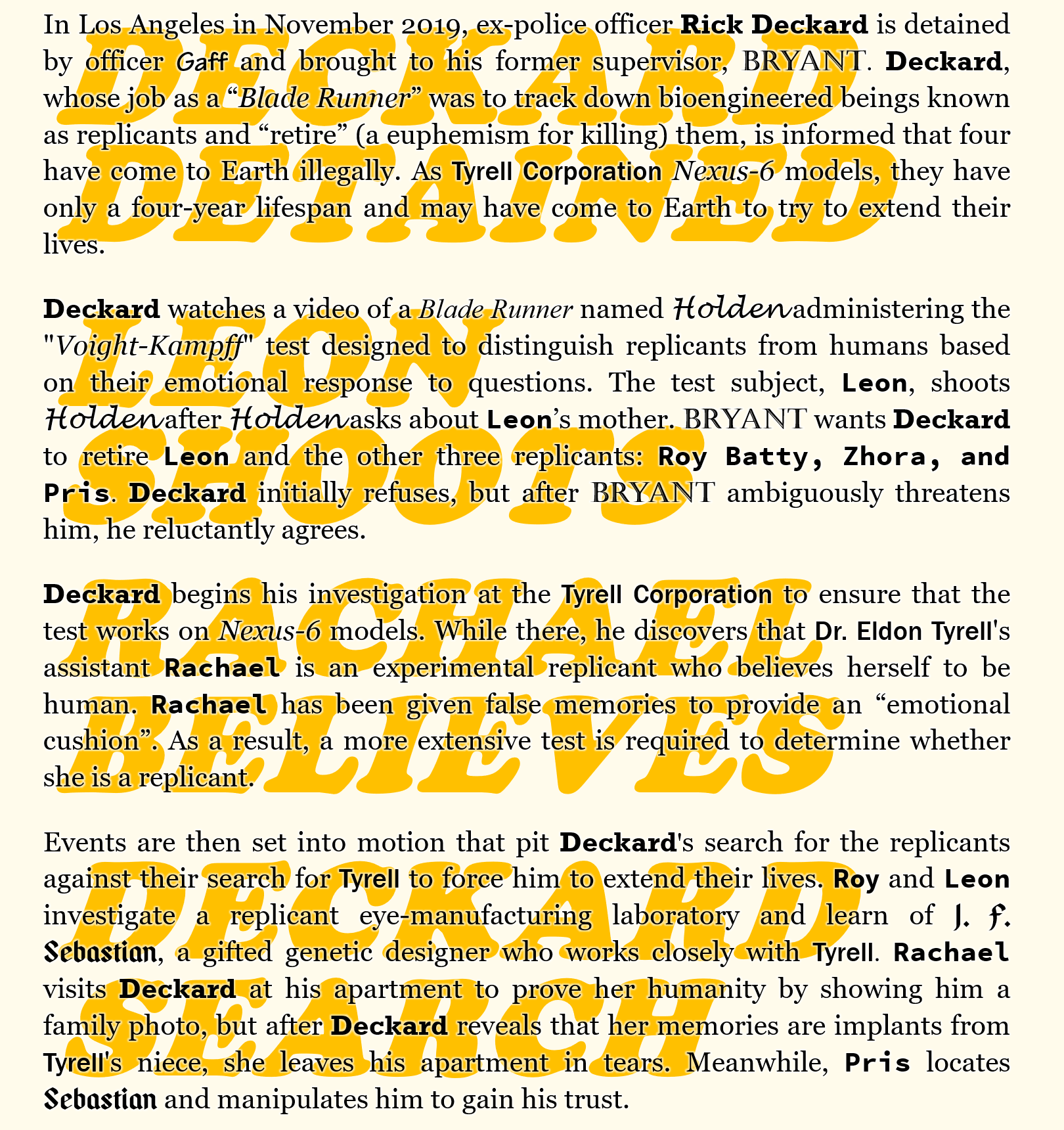}
 \caption{Body text over a two-word paragraph summary.}
 \label{fig:bladerunner}
\end{figure}

\autoref{fig:bladerunner} shows body text superimposed over summary words extracted via NLP. When reading a long text on a mobile device only a paragraph or two are visible, meaning that traditional textual cues such as headings and spacing may be off screen, thereby making it more difficult to navigate around the text. Using a large proper noun and associated verb could facilitate skimming while scrolling by providing textual landmarks to characters and their actions. This large text could disappear on scroll-stop. This is a summary operation (3.5) to reduce text, arguably to a high number of categories (3.1:D); and reducing cognitive load by not requiring the viewer to recall off-screen headings (3.2:D). 

\section{Conclusion}
The itemized literal benefits provide a starting point for researchers and designers to consider the alternative visualizations and the potential benefits across different design alternatives. These literal benefits can be considered in any kind of text-enhanced visualization, whether simple labels, or through to the examples shown here.

The example analyses here focus on collections of short semantic word sequences. The examples here may be highly-specific, however, the intent is to illustrate the possibilities and benefits of literal depictions. These should encourage digital humanities researchers to search for new visualization techniques. It should also encourage visualization researchers to validate benefits beyond prior research in adjacent fields through experimentation directly on text-orientated visualizations to quantify and characterize benefits. And, it should encourage linguistics and NLP researchers to innovate with humanities researchers to find and annotate word sequences, which can be used in “human-in-the-loop” analytical user interfaces for use by digital humanities scholars.

\acknowledgments{
Images created by the author are CC-BY-4.0. Some of these images and portions of the text will appear in the forthcoming book \textit{Visualizing with Text}, by Richard Brath, as part of the Visualization Series edited by Tamara Munzner to be published by A.K. Peters in November 2020. }


\bibliographystyle{abbrv-doi}

\bibliography{template}
\end{document}